\documentclass[prd,preprintnumbers,showpcs,amsmath,amssymb,nofootinbib, twocolumn,showpacs,notitlepage,superscriptaddress]{revtex4-1}
\usepackage{multirow}
\usepackage{epsfig}
\usepackage{amsmath}
\usepackage{bm}
\usepackage{times}
\usepackage{graphicx}
\usepackage{color}
\usepackage{slashed}
\usepackage{float}
\usepackage{diagbox}

\usepackage{xcolor}
\usepackage{graphicx}
\usepackage{amsmath}
\usepackage{hyperref}
\usepackage{soul}
\usepackage{lipsum}
\usepackage{ulem}
\usepackage{mathtools}
\usepackage{comment}
\usepackage{subcaption}
\usepackage{multirow}
\usepackage{booktabs}

\def\bea{\begin{eqnarray}}
\def\eea{\end{eqnarray}}
\def\bean{\begin{equation*}}
\def\eean{\end{equation*}} 

\def\beaal{\begin{align}}
\def\eeaal{\end{align}}
\definecolor{OliveGreen}{rgb}{0,0.6,0}

\newcommand{\Green}[1]{\textcolor{green!50!black}{#1}}
\newcommand{\Red}[1]{\textcolor{red!70!black}{#1}}

\begin{document}

\preprint{KCL-PH-TH-2025-44}

\title{A battle of designs: triangular vs. L-shaped detectors and parity-violation in the gravitational-wave background}

\author{Hannah Duval}
\email{hannah.marie.d.duval@vub.be}
\affiliation{Theoretische Natuurkunde, Vrije Universiteit Brussel, Pleinlaan 2, B-1050 Brussels, Belgium}

\author{Charles Badger}
\email{charles.badger@kcl.ac.uk}
\affiliation{Theoretical Particle Physics and Cosmology Group,  Physics Department, \\ King's College London, University of London, Strand, London WC2R 2LS, United Kingdom}

\author{Mairi Sakellariadou}
\email{mairi.sakellariadou@kcl.ac.uk}
\affiliation{Theoretical Particle Physics and Cosmology Group,  Physics Department, \\ King's College London, University of London, Strand, London WC2R 2LS, United Kingdom}

\author{Jacopo Uggeri}
\email{jacopo.uggeri@kcl.ac.uk}
\affiliation{Theoretical Particle Physics and Cosmology Group,  Physics Department, \\ King's College London, University of London, Strand, London WC2R 2LS, United Kingdom}

\date{\today}

\begin{abstract}
We investigate the prospects for detecting a parity-violating gravitational-wave background (GWB) with third-generation ground-based detector networks. We focus on a network consisting of one Einstein Telescope (ET) and two Cosmic Explorer (CE) detectors. In our analysis we vary the ET design, detector orientations, and arm lengths, in order to assess the impact of geometry and scale on detection capabilities. We find that parity-violation sensitivity is driven primarily by network geometry. In particular, detector orientation has a substantial influence on sensitivity to circular polarization. Given current observational constraints from the fourth observing run of the LIGO-Virgo-KAGRA Collaboration, we find that ET alone cannot confidently detect parity-violation in a flat GWB. 
\end{abstract}
\maketitle

\section{Introduction}
\label{Sec:Intro}
The superposition of independent gravitational waves (GW) of cosmological and astrophysical origin forms a gravitational-wave background (GWB).
The detection of such a background can shed light on early-universe mechanisms, such as inflation~\cite{Grishchuk:1974ny, Starobinsky:1979ty,Badger:2024ekb}, first-order phase transitions~\cite{Kamionkowski:1993fg, Apreda:2001us, Grojean:2006bp,Romero:2021kby,Badger:2022nwo}, and cosmic strings~\cite{Auclair:2019wcv,Damour:2001bk,LIGOScientific:2021nrg}, and aid in characterizing astrophysical compact binary coalescences (CBCs) sources and populations~\cite{KAGRA:2021duu}. Searches for an unpolarized GWB using data from the ground-based interferometers LIGO~\cite{TheLIGOScientific:2014jea} and Virgo~\cite{TheVirgo:2014hva} have placed upper limits on its energy density~\cite{LIGOScientific:2025bgj}. 

A multitude of mechanisms in the early universe can generate parity-violation (PV) \cite{Alexander_2006} that results in the production of asymmetric amounts of right- and left-handed circularly polarized GWs.  Parity-violating sources and their effects on the GWB have been studied in the literature, including those arising from the Chern-Simons modification to gravity~\cite{CS_PV, Bartolo_2021, Takahashi_2009} and from axion inflation \cite{AxInf_PV, Dimastrogiovanni_2013}, turbulence in the primordial plasma  \cite{Kamionkowski_1994, EdWitten_PhaseTrans, Hogan_PhaseTrans}, or by primordial magnetic fields coupled to the cosmological plasma
\cite{Brandenburg_1996,Christensson_2001,Kahniashvili_2010,Brandenburg_2019,Brandenburg:2021aln} . 

Probing for polarization in the GWB can place constraints on parity-violating theories~\cite{Crowder_2013}.
Parity-violating effects in the GWB have been studied in the context of current and future detector sensitivities~\cite{Seto_2007, LISA_PVCom, LISA_PV, Bartolo:2018qqn, Domcke_2020, Xu_2019, Martinovic:2021hzy, Badger:2021enh, Cruz:2024esk}, and have been constrained through analyses of LIGO--Virgo data~\cite{Crowder_2013, Martinovic:2021hzy, LIGOScientific:2025kry}.

While PV detection appears unlikely in current ground-based detectors, third-generation detectors show promise in detecting, or at least imposing tighter constraints on, circular polarization. 
As we show below, the choice of the detector network configuration will play a crucial role in the circular polarization detection prospects. 
In this study, we present a dual {\sl theoretical} (based on signal-to-noise ratio) and {\sl data-driven} (Bayesian) analysis for assessing the sensitivity of third-generation ground-based detector networks to parity-violating signatures in a general power-law GWB model.

\section{Formalism}
\label{Sec: formalism}
The right- and left-handed modes, defined through the circularly polarized bases $e^{\rm{R}} = (e^{+} + ie^{\times})/\sqrt{2}$ and $e^{\rm {L}} = (e^{+} - ie^{\times})/\sqrt{2}$ (with + and $\times$ the plus and cross polarizations, respectively) read
$h_{\rm R} = (h_{+} - ih_{\times})/\sqrt{2}$ and $h_{\rm L} = (h_{+} + ih_{\times})/\sqrt{2}$.

Right- and left-handed correlators are
\begin{equation}
\begin{aligned}
\begin{pmatrix}
\langle h_{R}(f,\hat{\Omega})
h_{R}^{*}(f',\hat{\Omega}') \rangle \\
\langle h_{L}(f,\hat{\Omega})
h_{L}^{*}(f',\hat{\Omega}') \rangle
\end{pmatrix}
&=
\begin{pmatrix}
I(f,\hat{\Omega}) + V(f,\hat{\Omega}) \\
I(f,\hat{\Omega}) - V(f,\hat{\Omega})
\end{pmatrix}
\\[-0.2em]
&\quad\times
\frac{\delta(f-f')\delta^2(\hat{\Omega}-\hat{\Omega}')}{4\pi}~,
\end{aligned}
\label{eq:RL_correlators}
\end{equation}
%
where $\langle\cdot\rangle$ is the ensemble average and $I, V$ are the Stokes parameters; $V$ characterizes the asymmetry between right- and left-handed polarized waves, and $I (\geq |V|)$ is the gravitational wave's total amplitude. 
The polarization degree $\Pi(f) = V(f)/I(f)\in [-1,1]$, with $\Pi=-1$ and $\Pi=+1$ defining fully left- and right-handed polarization, respectively, while $\Pi = 0$ stands for an unpolarized isotropic GWB. 
For the latter ($V = 0$), Eq.~(\ref{eq:RL_correlators}) reduces to the standard correlator for an unpolarized isotropic background.
The signal-to-noise ratios (SNRs) $\rho_{I}$ and $\rho_{V}$ for the $I$- and $V$-modes, respectively, 
of a circularly polarized GWB, are calculated for a detector network over an observing time $T$~\cite{Seto:2008sr}. 
The power-law integrated (PI) curve, $\Omega_{{\rm{PI}},I}$, characterizes a detector network’s $I$-mode sensitivity~\cite{Thrane:2013oya}. We extend the procedure to the $V$-mode to obtain the PI sensitivity curve, $\Omega_{{\rm{PI}},V}$, for the $V$-mode. Details of the underlying formalism are provided in Appendix \ref{app: I_V_det_sens}.
To search for parity-violating GWBs, we adapt the formalism described in~\cite{Seto:2008sr, PhysRevLett.99.121101} to third-generation ground-based detector networks.
\section{Parity-Violating Models}
\label{sec: models}
In the presence of a parity-violating GWB, the cross-correlation of the signals detected by a detector pair $d_1$ and $d_2$ is
\begin{equation}
    \langle s_{1}(f)s_{2}(f^\prime)\rangle =
    \delta(f - f^\prime)    \frac{3H_0^2}{8\pi f^3}
    \Omega^{\prime}_{\rm GW}(f)
    \gamma_I^{d_1d_2}(f)~,
\end{equation}
where $s_i \equiv h_{i} + n_{i}$ is the strain measured by detector $d_i$, consisting of the GW signal $h_i$ and detector noise $n_i$~\cite{PhysRevLett.99.121101, Crowder_2013} and we define an effective energy density
\begin{equation}
\label{eq:Omega_prime_def}
    \Omega_{\rm GW}^{\prime}(f) = \Omega_{\rm GW}(f) \Bigg[ 1+ \Pi(f)\frac{\gamma_V^{d_1 d_2}(f)}{\gamma_I^{d_1 d_2}(f)} \Bigg]~.
\end{equation}
Note that $\gamma_{I/V}^{d_1 d_2}$ are the normalized overlap reduction functions (ORFs) for the $I$- and $V$-modes 
defined as
\begin{equation}
    \gamma_I^{d_1 d_2}(f) = \frac{\beta_{12}}{8\pi}\int d\hat{\Omega} (F_{d_1}^+ F_{d_2}^{+*} + F_{d_1}^{\times} F_{d_2}^{\times*}) e^{2\pi if\hat{\Omega} \Delta\overrightarrow{x}}~,
    \label{eq:orf-I}
\end{equation}
\begin{equation}
    \gamma_V^{d_1 d_2}(f) = \frac{\beta_{12}}{8\pi i}\int d\hat{\Omega} (F_{d_1}^+ F_{d_2}^{\times *} - F_{d_1}^{\times} F_{d_2}^{+*}) e^{2\pi if\hat{\Omega} \Delta\overrightarrow{x}}~,
    \label{eq:orf-V}
\end{equation}
where $F^{A}_d$ is the beam pattern function for polarization $A=+,\times$. 
The factor $\beta_{12}$, defined from the opening angle $\theta_i$ of detector $d_i$, is
\begin{equation}
    \beta_{12} \equiv 5 (\sin{\theta_1}\sin{\theta_2})^{-1} .
    \label{eq:orf_normalization}
\end{equation}
This ensures $\gamma_{I}(0) = 1$ for co-located detectors with any opening angle\footnote{We note that sometimes in the literature the normalization choice will only ensure this condition for $90^{\circ}$ opening angles.}.
The ORFs can be calculated either analytically or numerically~\cite{Romano:2016dpx, Seto:2008sr, Caporali:2025dyf}. Here, we employ the analytical method, which agrees with the numerical results over the considered frequency range.
To model\footnote{The quantity $\Omega'_{\rm GW}(f)$ is an effective, detector-pair-dependent energy-density spectrum introduced so that the cross-correlation retains the same form as in the unpolarized case; it should not be confused with the physical GWB spectrum $\Omega_{\rm GW}(f)$.} $\Omega_{\rm GW}^{\prime}(f)$, we assume that the underlying GWB energy-density spectrum follows a power law 
\begin{equation}
    \Omega_{\rm{GW}}(f) = \Omega_{\alpha}(f/f_{\rm{ref}})^\alpha ,
    \label{eq:omega_power_law_model}
\end{equation}
where $\Omega_{\alpha}$ is the amplitude at the reference frequency ($f_{\rm ref} = 25~\rm{Hz}$), and $\alpha$ is the spectral index~\cite{Martinovic:2021hzy}. 
Theoretically motivated polarized GWBs can be modeled by specific choices of $\alpha$.
In this parametrization, the polarization fraction $\Pi(f)$ encodes any difference in spectral behavior between the unpolarized $I(f)$ and parity-violating $V(f)$ components of the GWB.
For simplicity, we assume a frequency-independent polarization fraction, $\Pi(f)=\Pi$, so that the intensity and circular-polarization  components of the GWB share the same power-law frequency dependence, but the formalism can be straightforwardly generalized to arbitrary polarized GWBs through a frequency-dependent $\Pi(f)$~\cite{Martinovic:2021hzy}. This choice is well motivated by several theoretical parity-violating GWB models, such as those sourced by non-Abelian axion inflation~\cite{Badger:2024ekb, Maleknejad:2016qjz}, string-inspired cosmological models~\cite{CS_PV}, and some turbulence scenarios~\cite{Kahniashvili_2005}. 
Nevertheless, the effective quantity $\Omega'_{\rm GW}(f)$ entering the detector cross-correlation is generally not an exact power law because it is defined through the frequency-dependent ratio $\gamma_V^{d_1d_2}(f)/\gamma_I^{d_1d_2}(f)$.


\section{ET-CE detector network configurations}
\label{Sec: 3Gdet}
We consider seven representative networks, each combining one Einstein Telescope (ET) with two Cosmic Explorer (CE) detectors, using the geometric information from~\cite{Ebersold:2024hgp}.
We include triangular ET designs with $10$ and $15$ km arms ($\Delta_{10}$, $\Delta_{15}$), as well as ET designs consisting of two L-shaped detectors with $15$ km arms, denoted S (Sardinia) and R (Meuse-Rhine region) 
following the standard ET configurations introduced in~\cite{Branchesi:2023mws} and extended to ET + CE networks in~\cite{Ebersold:2024hgp}. 
For CE, we consider L-shaped detectors with $40$ km (P) and $20$ km (Y) arms, corresponding to the Pacific Ocean and Gulf of Mexico site configurations adopted in~\cite{Gupta:2023lga, Ebersold:2024hgp}.
The triangular configurations consist of three detectors, while an L-shaped configuration consists of one. Thus, a $\Delta$ + (PY) network contains five detectors ($10$ baselines), while an (SR) + (PY) network contains four ($6$ baselines).
These configurations explore how detector orientation and arm length influence network performance, with some optimized for GWB sensitivity and others for CBC source localization, clarifying whether differences between triangular and L-shaped designs arise from geometry or orientation. 
The full set of configuration names and descriptions is in Table~\ref{table:configs}.
Detector specifications are listed in Appendix \ref{app:technicalities}, 
based on \cite{Ebersold:2024hgp}. We use the amplitude spectral density curves found in \cite{ETdesign} and \cite{CEdesign}.

\begin{table}[h!]
\centering
\begin{tabular}{|c|c|} 
 \hline
 Network name & Description \\ 
 \hline
 $\Delta_{10}$(PY)$_4$ 
 & $10$ km $\Delta$ + 2CE, GWB-optimized \\ 
 \hline
$\Delta_{10}$(PY)$_5$ 
 & $10$ km $\Delta$ + 2CE, Localization-optimized \\
 \hline
 $\Delta_{15}$(PY)$_4$ 
 & $15$ km 
 $\Delta$ + 2CE, GWB-optimized \\ 
 \hline
 $\Delta_{15}$(PY)$_5$  
 & $15$ km $\Delta$ + 2CE, Localization-optimized \\
 \hline
 (SR)$_1$(PY)$_1$ 
 & $15$ km  2L + 2CE, GWB-optimized \\
 \hline
 (SR)$_2$(PY)$_2$  
 & $15$ km 2L + 2CE, Localization-optimized \\
 \hline
 (SR)$_3$(PY)$_3$  
 & $15$ km 2L + 2CE, Hybrid \\
 \hline
\end{tabular}
\caption{The different third-generation detector network configurations that are considered in this study, which follow the definitions and optimization procedures of \cite{Ebersold:2024hgp}. “GWB-optimized” denotes  configurations favoring GWB sensitivity,
constructed by minimizing their PI-curve minimum, while “Localization-optimized” configurations favor source localization and are computed to maximize the network-alignment factor, which measures the relative sensitivity to the cross polarization \cite{Klimenko:2005xv}. “Hybrid” configurations provide a balanced design providing a trade-off between the two.}
\label{table:configs}
\end{table}

\begin{figure}[htbp]
    \centering
    \includegraphics[width=\linewidth]{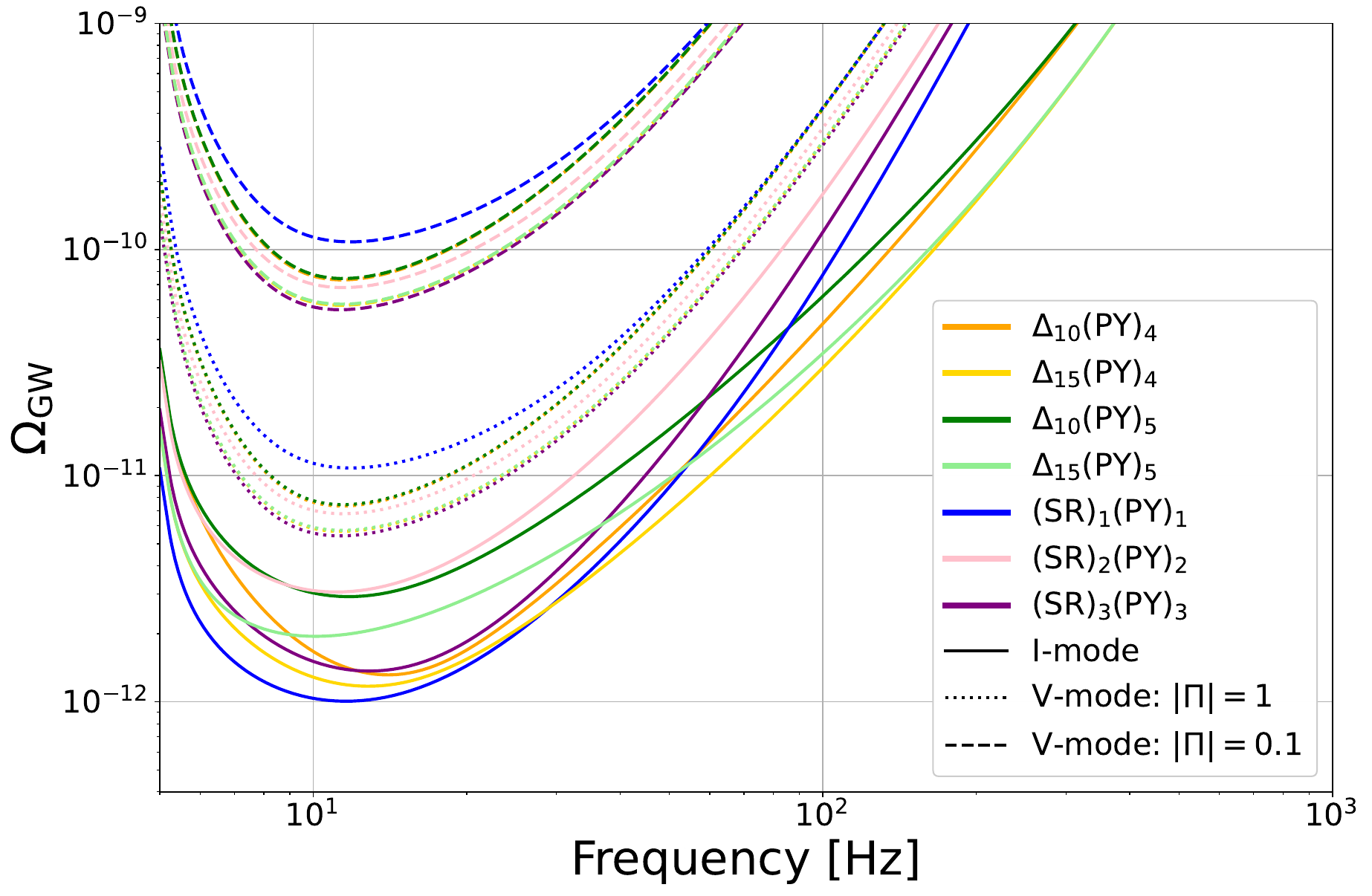}
    \caption{PI sensitivity curves for $I$-mode (solid lines) and $V$-mode (dotted and dashed lines), shown for the seven different detector third-generation network configurations for an observing time of one year and $\rho=6$. The $V$-mode curves correspond to $|\Pi| = 1$ (dotted) and $|\Pi| = 0.1$ (dashed).}
    \label{fig:PI_curves}
\end{figure}

In Fig. \ref{fig:PI_curves}, we present the $I$- and $V$-mode PI curves for the various network configurations, extending previous ET design studies \cite{Branchesi:2023mws} to networks combining ET with CE and to $V$-mode sensitivity. Further details can be found in Appendix \ref{app: I_V_det_sens}.
The $\Omega_{{\rm{PI}},I}$ curves show that GWB-optimized networks yield the highest $I$-mode sensitivity, which is expected as they were specifically designed to maximize sensitivity to GWBs.
In particular, (SR)$_1$(PY)$_1$ dominates below 20 Hz, while at higher frequencies all triangular configurations outperform the 2L designs.

The $\Omega_{{\rm{PI}},V}$ curves exhibit reduced sensitivity compared to their $I$-mode counterparts, with sensitivity decreasing as $|\Pi|$ approaches 0.
Their ranking is relatively stable across frequencies: (SR)$_3$(PY)$_3$ is consistently the most sensitive configuration, with $\Delta_{15}$(PY)$_4$ and $\Delta_{15}$(PY)$_5$ following as close seconds, while (SR)$_1$(PY)$_1$ performs the worst. Overall, the 2L network designs exhibit the largest variability in performance, reflecting their strong dependence on relative orientation of the detectors. In contrast, networks involving the triangular ET configuration show a much more stable performance across different realizations.
In particular, the superior performance for $V$-modes of the hybrid network (SR)$_3$(PY)$_3$ compared to the fully aligned (SR)$_1$(PY)$_1$, which excel at $I$-mode detection, indicates that 2L ET networks are especially sensitive to relative orientation.
We also present parity-violation detection prospects for ET alone in Appendix~\ref{app:ET_alone}, confirming that current constraints~\cite{LIGOScientific:2025kry} have already excluded the parameter space in which ET alone would be capable of detecting circular polarization. Therefore, the inclusion of CE detectors is a necessity to make robust parity-violation inferences.

\section{Detection prospects}
\label{sec: results}
We present {\sl theoretical} (based on SNR) and {\sl data-driven} (obtained with a Bayesian analysis) prospects for detecting PV by applying the previously described formalisms to third-generation detector networks.
Our analysis covers general power-law GWBs, discussed in Section~\ref{sec: models}, with $-12.5 \leq \log_{10}\Omega_{\alpha} \leq -10$, spectral indices $\alpha = -3, \dots, 3$, and polarization degrees $-1 \leq \Pi \leq 1$.
We note that the upper bound on the $\log_{10}\Omega_{\alpha}$ range is determined by current detector constraints~\cite{LIGOScientific:2025bgj}, while the lower bound is to allow studied GWBs to be potentially detected in the considered networks, as seen in Fig. \ref{fig:PI_curves}. 

\subsection{Signal-to-noise ratio}
\label{sec:snr-prospects}
\begin{figure}[htbp]
    \centering
    \includegraphics[width=\linewidth]{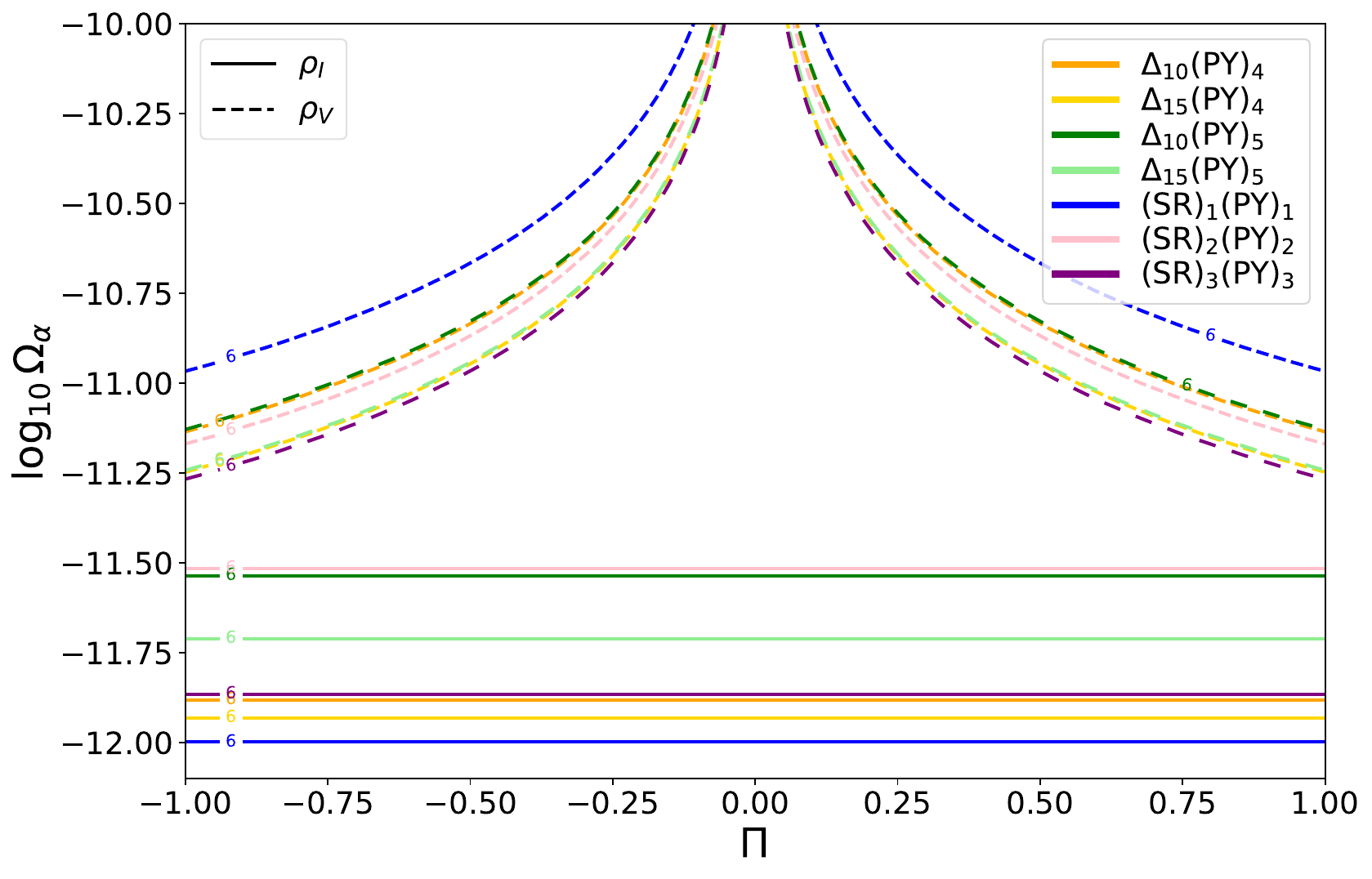}
    \caption{SNRs for $I$-mode (solid lines) and $V$-mode (dashed lines) for the seven third-generation detector network configurations, assuming a flat ($\alpha=0$) power-law GWB.}
    \label{fig: SNR_theory}
\end{figure}

The sensitivities of the third-generation detector networks are quantified by the $I$- and $V$-mode SNRs. Fig.~\ref{fig: SNR_theory} shows the results for a flat PL ($\alpha=0$) spectrum across the seven network configurations. 
As expected, $\rho_I$ is independent of $\Pi$ and scales only with the GWB amplitude, while $\rho_V$ grows with $|\Pi|$, enhancing detectability for more strongly polarized backgrounds. 
We repeat this study for other power-law indices $\alpha=-3, ... 3$ and rank their detection prospects in Tables~\ref{tab:SNR_I_rankings} and~\ref{tab:SNR_V_rankings}.

\begin{table*}[t]
\centering

\small
\begin{tabular}{|l|c|c|c|c|c|}
\hline
\diagbox
  {Rank}
  {\raisebox{-1.5ex}{$\alpha$}} 
& $-3,-2$ 
& $-1$
& $0, 1$
& $2$ 
& $3$ \\
\hline

1
& (SR)$_1$(PY)$_1$ 
& (SR)$_1$(PY)$_1$ 
& (SR)$_1$(PY)$_1$ 
& $\Delta_{15}$(PY)$_4$ 
& $\Delta_{15}$(PY)$_4$\\

\hline
2
& $\Delta_{15}$(PY)$_4$ 
& $\Delta_{15}$(PY)$_4$ 
& $\Delta_{15}$(PY)$_4$ 
& (SR)$_1$(PY)$_1$ 
& $\Delta_{15}$(PY)$_5$ \\

\hline
3
& $\Delta_{15}$(PY)$_5$ 
& (SR)$_3$(PY)$_3$ 
& $\Delta_{10}$(PY)$_4$
& $\Delta_{15}$(PY)$_5$ 
& $\Delta_{10}$(PY)$_4$ \\

\hline
4
& (SR)$_3$(PY)$_3$ 
& $\Delta_{10}$(PY)$_4$ 
& (SR)$_3$(PY)$_3$
& $\Delta_{10}$(PY)$_4$ 
& $\Delta_{10}$(PY)$_5$ \\

\hline
5
& $\Delta_{10}$(PY)$_4$ 
& $\Delta_{15}$(PY)$_5$ 
& $\Delta_{15}$(PY)$_5$ 
& (SR)$_3$(PY)$_3$ 
& (SR)$_1$(PY)$_1$ \\

\hline
6
& (SR)$_2$(PY)$_2$ 
& (SR)$_2$(PY)$_2$ 
& $\Delta_{10}$(PY)$_5$ 
& $\Delta_{10}$(PY)$_5$ 
& (SR)$_3$(PY)$_3$ \\

\hline
7
& $\Delta_{10}$(PY)$_5$ 
& $\Delta_{10}$(PY)$_5$ 
& (SR)$_2$(PY)$_2$ 
& (SR)$_2$(PY)$_2$ 
& (SR)$_2$(PY)$_2$ \\

\hline
\end{tabular}
\caption{Ranking from best (1st) to worst (7th) of detector networks by $I$-mode SNRs for integer PL indices in the range~$\alpha \in [-3, 3]$, grouping together indices that yield identical prospects. We also note that the physically motivated value of $\alpha = 2/3$ yields the same ranking as the central $0, 1$ column.}
\label{tab:SNR_I_rankings}
\end{table*}

\begin{table}[h!]
\centering
\begin{tabular}{|l|c|}
\hline
\diagbox
  {Rank}
  {\raisebox{-1.5ex}{$\alpha$}}  & $[-3, 3]$ \\
\hline
1 & (SR)$_3$(PY)$_3$ \\
\hline
2 & $\Delta_{15}$(PY)$_4$ \\
\hline
3 & $\Delta_{15}$(PY)$_5$ \\
\hline
4 & (SR)$_2$(PY)$_2$ \\
\hline
5 & $\Delta_{10}$(PY)$_4$ \\
\hline
6 & $\Delta_{10}$(PY)$_5$ \\
\hline
7 & (SR)$_1$(PY)$_1$ \\
\hline
\end{tabular}
\caption{Ranking from best (1st) to worst (7th) of detector networks by $V$-mode SNRs for different PL indices $\alpha$.}
\label{tab:SNR_V_rankings}
\end{table}

We find that the 2L network (SR)$_1$(PY)$_1$ provides the largest $\rho_I$ for $-3 \leq \alpha \leq 1$, while
the $15$ km triangular network $\Delta_{15}$(PY)$_4$ overtakes it at $\alpha = 2$, highlighting the advantage of a triangular design for GWBs with large positive spectral indices.
For $V$-modes, we find a stable ranking throughout the considered $\alpha$ range, which is consistent with the ordering seen in the PI curves in Fig.~\ref{fig:PI_curves} whose frequency-dependent shapes qualitatively explain the SNR rankings across different spectral indices.
We highlight that fixing different reference frequencies result in the same ordering of SNR prospects, emphasizing the generality of our results.
Appendix~\ref{app:ET_alone} presents ET-only configurations. These results show that all triangular ETs suffer a dramatic loss of PV sensitivity compared to 2L designs.

\subsection{Bayesian analysis}
We now complement the SNR analysis with a data-driven Bayesian analysis on simulated signals, assuming the design sensitivity and noise characteristics of each third-generation network configuration.
Parameter estimation is carried out with the  {\tt Bilby}~\cite{Ashton:2018jfp} package using the {\tt Dynesty} sampler~\cite{Speagle:2019ivv}. 
This allows us to assess a network's capability to detect a polarized signal and to distinguish it from an unpolarized one. 
To this end, we fix a power-law index $\alpha$ and generate 1600 simulated injections uniformly over the $\log_{10}\Omega_{\alpha}$ and $\Pi$ range in each network configuration, then searching over the priors of Table~\ref{table:priors} based off of theoretical bounds and previously explored search priors~\cite{Martinovic:2020hru, LIGOScientific:2025kry}.
The procedure is then repeated with $\Pi = 0$ to model unpolarized backgrounds, using the same amplitude and spectral index priors. 
To reduce statistical noise, we generate $30$ independent injection sets with random detector noise and combine results with an optimized weighted average~\cite{Romano:2016dpx}.
This enables a direct comparison between polarized and unpolarized searches. 
\begin{table}[h!]
\centering
\begin{tabular}{|c|c|} 
 \hline
 Parameters $\theta$ & Priors \\ 
 \hline
 $\log_{10}\Omega_{\rm \alpha}$  & \rm Uniform($-18, -7$) \\ 
 \hline
 $\alpha$ & \rm Gaussian($0, 3.5$) \\
 \hline
 $\Pi$ & \rm Uniform($-1, 1$) \\
 \hline
\end{tabular}
\caption{Priors used in the Bayesian analysis. The $\log_{10} \Omega_{\rm \alpha}$ and $\alpha$ priors follow previous searches, while taking into account the sensitivity of third-generation detectors~\cite{PhysRevD.104.022004, Martinovic:2020hru}. The prior on $\Pi$ is the full, physically allowed range of values~\cite{Martinovic:2021hzy}.
}
\label{table:priors}
\end{table}

\subsubsection{Comparison to signal-to-noise}
\begin{figure}[h!]
    \centering
    \includegraphics[width=\linewidth]{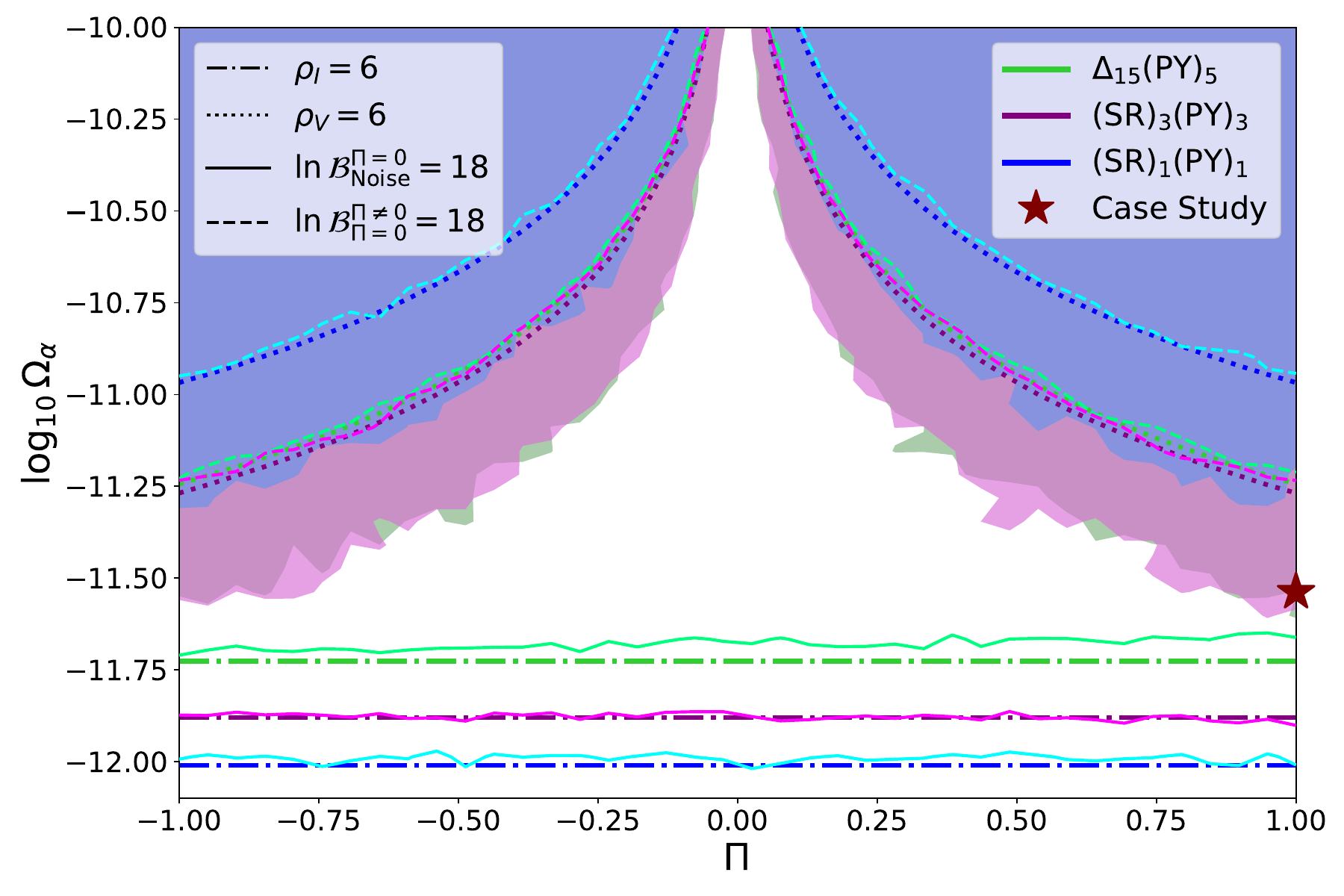}
    \caption{SNR and log Bayes detection prospects for $I$- and $V$-mode in a flat GWB in the $\Delta_{10}$(PY)$_5$, (SR)$_3$(PY)$_3$ and (SR)$_1$(PY)$_1$ configurations. 
    The green, purple and blue filled regions correspond to $95 \%$ exclusion regions of $\Pi=0$.
    The dark red star marks the case study discussed in Sec.~\ref{subsec: dataAnalysis_PE}.}
    \label{fig: Theory_vs_Data_Compare}
\end{figure}

We compare the recovered data-analysis Bayes factors to the $I$- and $V$-mode SNRs for all injections.
Fig.~\ref{fig: Theory_vs_Data_Compare}
shows this comparison for the $\Delta_{15}$(PY)$_4$, (SR)$_3$(PY)$_3$ and $\Delta_{10}$(PY)$_5$ configurations for a flat GWB spectrum.
More concretely, it shows the SNRs alongside the log Bayes factors for 
an unpolarized model compared to detector noise, $\ln\mathcal{B}_{\rm Noise}^{\Pi = 0}$, and for a polarized model relative to an unpolarized one, $\ln\mathcal{B}_{\Pi = 0}^{\Pi \neq 0}$. 
We adopt strong detection thresholds of $\rho_I=6$ and $\rho_V=6$ \cite{Romano:2016dpx}. 
Using the equivalence $\rho^2 \simeq 2 \ln\mathcal{B}$ in the strong signal regime~\cite{Romano:2016dpx}, these thresholds correspond to $\ln\mathcal{B}_{\rm Noise}^{\Pi = 0}=18$ and $ \ln\mathcal{B}_{\Pi = 0}^{\Pi \neq 0}=18$.
Our findings confirm the validity of this relation.
In conclusion, the SNR predictions reproduce the trends seen in the data-driven analysis across the three networks, demonstrating that simple theoretical estimates based on SNRs provide an excellent guide to the detectability of polarized GWBs.
We note that the present analysis assumes uncorrelated detector noise. For the ET triangular configuration, correlated Newtonian noise between the co-located detectors is expected to further degrade the sensitivity to GWBs \cite{Janssens:2022xmo, Janssens:2024jln, Caporali:2025mum}. 
Since such correlated noise is expected to be suppressed over the large separations between L-shaped detectors, it would primarily degrade the performance of networks containing triangular detectors.

\subsubsection{Parameter estimation}
\label{subsec: dataAnalysis_PE}
We perform a Bayesian analysis to assess parameter recovery of simulated parity-violating GWBs for the $\Delta_{15}$(PY)$_5$, (SR)$_3$(PY)$_3$ and (SR)$_1$(PY)$_1$ configurations.
From the marginalized posteriors, we compute the $95\%$  
credible intervals on $\Pi$, allowing us to make $\Pi = 0$ exclusion analyses.
This is illustrated in Fig.~\ref{fig: Theory_vs_Data_Compare}, where the green, purple and blue shaded regions indicate parameter space in which $\Pi=0$ can be excluded at $95\%$ confidence by the $\Delta_{15}$(PY)$_5$, (SR)$_3$(PY)$_3$ and (SR)$_1$(PY)$_1$ configurations, respectively. 

\begin{figure}[h!]
    \centering
    \includegraphics[width=0.48\textwidth]{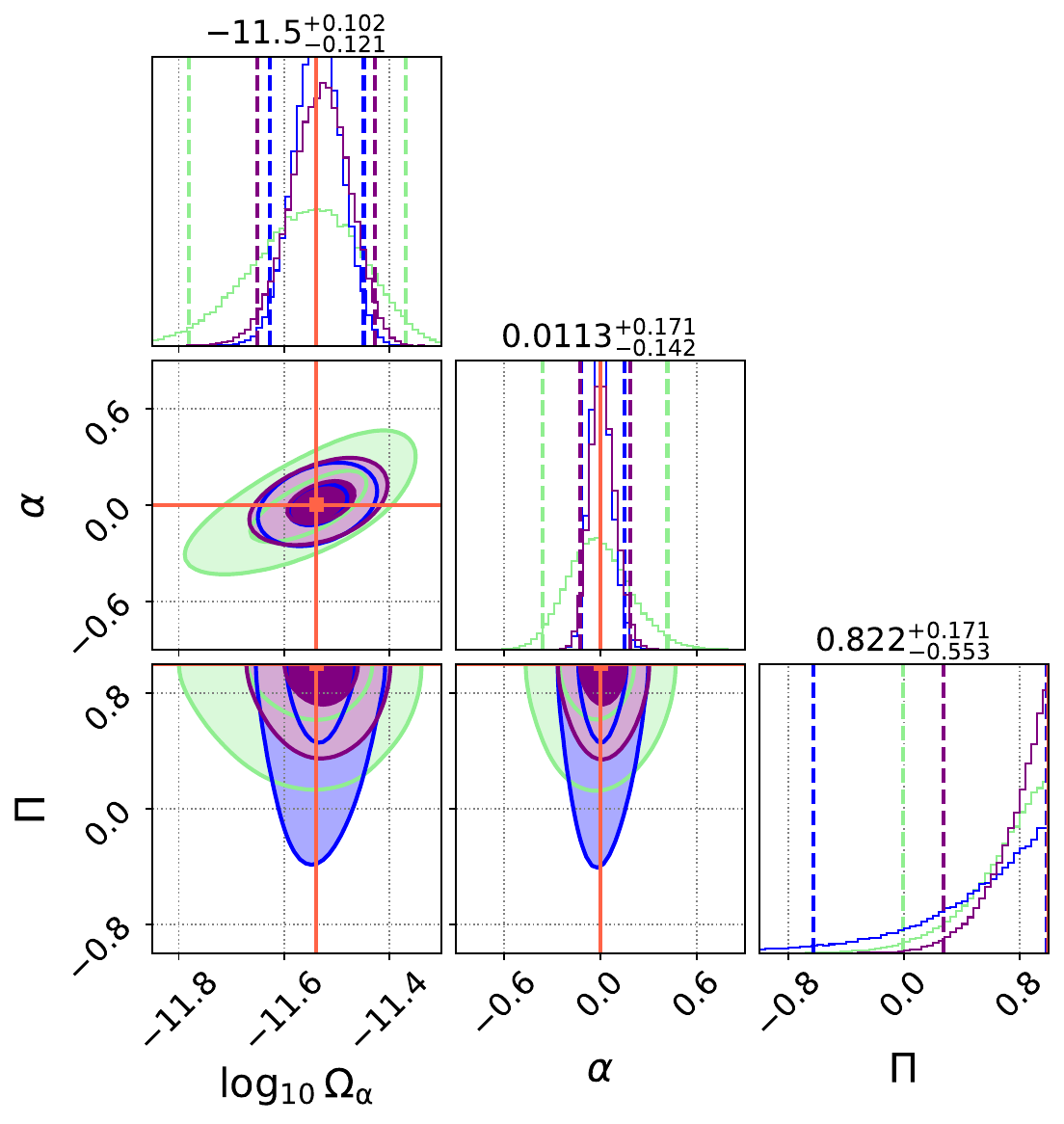}
    \caption{Posterior distributions from Bayesian inference for parity-violating power-law GWBs, combined over $30$ simulated runs for $f_{\rm ref}=25$ Hz. Shown are the $\Delta_{15}$(PY)$_5$ (green), (SR)$_3$(PY)$_3$ (purple) and (SR)$_1$(PY)$_1$ (blue) networks, with dark and light shading indicating $1\sigma$ and $2\sigma$ credible regions.  
    Here: $\Pi = 1$ and $\log_{10} \Omega_{\alpha} = -11.54$ shown in red.}
    \label{fig:PE}
\end{figure}

Next, we focus on a flat parity-violating GWB case study with $\log_{10} \Omega_{\alpha}=-11.54$, $\Pi=1$, indicated by the red star in Fig.~\ref{fig: Theory_vs_Data_Compare}. These parameters were chosen such that the signal lies within the exclusion regions of the $\Delta_{15}$(PY)$_5$ and (SR)$_3$(PY)$_3$ configurations, but remains outside that of the (SR)$_1$(PY)$_1$ network, thereby probing the threshold for excluding $\Pi=0$ across the different detector configurations.
The combined posterior distributions recovered for this benchmark are shown in Fig.~\ref{fig:PE}, where the injected model parameters are recovered within $1\sigma$ for all three networks.
We derive the $95\%$ and $99\%$ lower limits (LL) on $\Pi$ and compute the log Bayes factors for the parity-violating and the unpolarized GWB hypotheses. The results are summarized in Table~\ref{table:PE_results}. 
\begin{table}[h!]
\centering

\begin{tabular}{|c|c|c|c|} 
 \hline
  & $\Delta_{15}$(PY)$_5$ & (SR)$_3$(PY)$_3$ & (SR)$_1$(PY)$_1$ \\ 
 \hline
 $95 \% ~\text{LL } \Pi$ & $\Green{0.130}$ & $\Green{0.374}$ & $\Red{-0.399}$ \\
 \hline
 $99 \% ~\text{LL } \Pi$ & $\Red{-0.191}$ & $\Green{0.142}$ & $\Red{-0.822}$ \\
 \hline
 $\ln\mathcal{B}_{\rm Noise}^{\Pi= 0}$  & $35.377 \pm 0.019$ & $109.361 \pm 0.021$ & $\mathbf{195.291 \pm 0.021}$ \\
 \hline
 $\ln\mathcal{B}_{\Pi=0}^{\Pi\neq 0}$ &$2.570 \pm 0.028$ & $\mathbf{4.501 \pm  0.031}$ & $0.680 \pm   0.031$ \\
 \hline
\end{tabular}
\caption{$95\%$ and $99\%$ LL of $\Pi$ for the three networks, together with the log Bayes factors comparing the parity-violating GWB signal hypothesis to the noise-only hypothesis for the considered case study. For legibility, we highlight in green and red values that show respectively whether the network can exclude $\Pi = 0$ or not, at the required confidence interval for each row. We also highlight the highest log Bayes values per row in bold.}
\label{table:PE_results}
\end{table}

The log Bayes factors are consistent with our theoretical results in Section~\ref{sec:snr-prospects}. All network configurations robustly identify the presence of an unpolarized GWB, with (SR)$_1$(PY)$_1$ being the most decisive, as expected from its optimization for $I$-mode detection. The $\ln\mathcal{B}_{\Pi=0}^{\Pi\neq 0}$ values also exhibit a clear ranking: (SR)$_3$(PY)$_3$ provides strong evidence for a parity-violating GWB, followed by $\Delta_{10}$(PY)$_5$, while (SR)$_1$(PY)$_1$ yields inconclusive results.
The LL constraints on $\Pi$ offer a more directly interpretable summary: (SR)$_3$(PY)$_3$ rules out $\Pi = 0$ at both $99\%$ and $95\%$ confidence, $\Delta_{10}$(PY)$_5$ excludes $\Pi = 0$ at $95\%$ confidence only, and (SR)$_1$(PY)$_1$ does not exclude $\Pi = 0$ at either level.

In conclusion, these findings highlight that network design strongly impacts parameter constraints, particularly for circular polarization $\Pi$, with 
(SR)$_3$(PY)$_3$ providing the tightest constraints on parity-violating GWBs and (SR)$_1$(PY)$_1$ the weakest.

\section{Conclusions}
\label{sec: conclusions}
In this study, the detection prospects of a general parity-violating GWB with third-generation detector networks were investigated using a SNR and a Bayesian analysis approach.
By extending the standard GWB formalism to include the $V$-mode, we introduced PI curves and SNR prospects that accurately reproduce data-based results, providing an efficient proxy for assessing circular polarization detection prospects.

We considered an ET detector, in both triangular and 2L designs, paired with two CE detectors across a range of orientations, and ET arm lengths.
A broad set of power-law GWB spectra, motivated by cosmological parity-violating mechanisms, was analyzed.

We find that both detector geometry and scale play important roles in determining parity-violation sensitivity. Among the triangular ET configurations, the $15$ km designs consistently outperform their $10$ km counterparts.
We find that the weakest and strongest networks differ only in their alignment angles. The (SR)$_1$(PY)$_1$ configuration, which is optimized for $I$-mode sensitivity, exhibits the weakest constraints on $\Pi$ among the networks considered, while (SR)$_3$(PY)$_3$, defined as a hybrid compromise between GWB detection and source localization, provides the strongest constraining power over $\Pi$. 
This indicates that optimizing a network for a balance of localization and GWB performance need not come at the expense of parity-violation sensitivity and may, in fact, improve it.
Parameter estimation studies further demonstrate that network design strongly affects the ability to constrain the GWB parameters.
We find that more sensitive configurations enable the exclusion of $\Pi=0$ at $95\%$ 
in a noticeably wider region of parameter space than less sensitive detector designs. 
In Appendix~\ref{app:ET_alone}, we show that PV detection in ET alone has been ruled out; circular polarization in a third-generation network requires CE detectors. These findings underscore the importance of global network design choices in order to optimize the discovery potential for fundamental physics in the GWB. 
These results highlight how next-generation detectors could open a direct observational window onto fundamental parity-violating physics in the early Universe.

\begin{acknowledgments}
We are especially grateful to Michael Ebersold for valuable insights on third-generation network configurations. 
We also thank Alberto Mariotti and Miguel Vanvlasselaer for helpful discussions, and Gianmassimo Tasinato and Angelo Ricciardone for useful comments.
This document has an ET document number ET-0483A-25. The work of MS is partially supported by the Science and Technology Facilities Council (STFC grant ST/X000753/1).
HD is supported by the Strategic
Research Program High-Energy Physics of the Research
Council of the Vrije Universiteit Brussel, the iBOF “Unlocking the Dark Universe with Gravitational Wave Observations: from Quantum Optics to Quantum Gravity”
of the Vlaamse Interuniversitaire Raad. JU holds a KCL STFC-funded PhD position.
\end{acknowledgments}

\newpage

\clearpage
\appendix

\section{\texorpdfstring{$I$- and $V$-mode detector sensitivity}{I- and V-mode detector sensitivity}}\label{app: I_V_det_sens}
The total intensity of the GWB is described by the Stokes parameter $I(f)$, which is related to the GW energy density spectrum through 
\begin{equation}
    I(f) \equiv \frac{3 H_0^2 \Omega_{\rm GW}(f)}{8 \pi^2 f^3}.
\end{equation}
The circularly polarized component of the background is described by the Stokes parameter $V(f)$.
Defining $\Pi(f)$ as the degree of circular polarization, we write
\begin{equation}
    \Pi(f) \equiv \frac{V(f)}{I(f)},
\end{equation}
or equivalently,
\begin{equation}
    V(f) \equiv \frac{3 H_0^2 \Pi(f) \Omega_{\rm GW}(f)}{8 \pi^2 f^3 } .
\end{equation}
Using~\cite{Seto:2008sr}, the SNR for mode $A\in\{I,V\}$ is
\begin{equation}
\rho_A
=
\sqrt{2T}
\left[
\int_{f_{\rm min}}^{f_{\rm max}} df\,
\frac{\Omega_{\rm GW}^2(f)}
{\Omega_{{\rm eff},A}^2(f)}
\right]^{1/2}.
\label{eq:SNR_tot}
\end{equation}
The effective energy densities are
\begin{align}
\Omega_{{\rm eff},I}(f)
&=
\frac{2\pi^2}{3H_0^2} f^3
\left[
\frac{\Gamma_{\rm eff}^{VV}}
{\Gamma_{\rm eff}^{II}\Gamma_{\rm eff}^{VV}
-\left(\Gamma_{\rm eff}^{IV}\right)^2}
\right]^{1/2},
\label{eq:Omega_eff_I}
\\
\Omega_{{\rm eff},V}(f)
&=
\frac{2\pi^2}{3H_0^2} f^3
\left[
\frac{\Gamma_{\rm eff}^{II}/\Pi^2(f)}
{\Gamma_{\rm eff}^{II}\Gamma_{\rm eff}^{VV}
-\left(\Gamma_{\rm eff}^{IV}\right)^2}
\right]^{1/2}.
\label{eq:Omega_eff_V}
\end{align}
Here all $\Gamma_{\rm eff}^{AB}$ are evaluated at frequency $f$, with
\begin{equation}
\Gamma_{\rm eff}^{AB}(f)
=
\sum_{i=1}\sum_{j>i}
\frac{
\Gamma_A^{d_id_j}(f)\Gamma_B^{d_id_j}(f)
}{
P_{n,i}(f)P_{n,j}(f)
},
\label{eq:Gamma_eff}
\end{equation}
and
\begin{equation}
\Gamma_A^{d_1d_2}(f)
=
\beta_{12}^{-1}\gamma_A^{d_1d_2}(f).
\label{eq:Gamma_beta}
\end{equation}
where 
$P_{n,i}$ is the noise power spectral density of detector $d_i$ and we use $\beta_{ij}$ as defined in Eq.~\eqref{eq:orf_normalization} and $\gamma_A^{d_1d_2}(f)$ as defined in Eqs.~\eqref{eq:orf-I} and~\eqref{eq:orf-V}. 
Note that as $\Pi$ approaches 0, $\rho_V$ also approaches 0 and the $\Omega_{{\rm{eff}}, V}$ curve grows infinitely large - both behaviors capturing the increasingly more difficult prospects of $V$-mode detection as it decreases.

\section{Technical information detector configurations}\label{app:technicalities}
In this Appendix we give the detailed information for our detector configurations. In Table \ref{tab:network_coords}, the latitude and longitude of each network center, the X-and Y-arm endpoints are given as well as the azimuths of the X-and Y-arms.

\begin{table*}[htbp]
\centering
\footnotesize
\setlength{\tabcolsep}{3pt}
\begin{tabular}{|c|c|c|c|c|}
 \hline
 \textbf{Name} & \textbf{Origin (lat, lon)} & \textbf{X-arm end (lat, lon)} & \textbf{Y-arm end (lat, lon)} & \textbf{Azimuth [°] X-/Y-arm} \\
 \hline
 E1 & (43.6300, 10.5000) & (43.7149, 10.5413) & (43.6983, 10.4193) & 19.43 / 319.43 \\ 
 E2 & (43.7200, 10.5500) & (43.7034, 10.4280) & (43.6351, 10.5088) & 259.43 / 199.43 \\
 E3 & (43.7000, 10.4200) & (43.6316, 10.5006) & (43.7164, 10.5420) & 139.43 / 79.43 \\
 \hline
 S1 & (40.5200, 9.4170) & (40.4950, 9.5909) & (40.6528, 9.4496) & 100.60 / 10.60 \\
 S2 & (40.5200, 9.4170) & (40.5699, 9.5815) & (40.6454, 9.3513) & 68.25 / 338.25 \\
 S3 & (40.5200, 9.4170) & (40.5976, 9.5620) & (40.6305, 9.3151) & 54.90 / 324.90 \\
 \hline
 R1 & (50.7200, 5.9210) & (50.5865, 5.8910) & (50.7007, 6.1312) & 188.14 / 98.14 \\
 R2 & (50.7200, 5.9210) & (50.5878, 5.8794) & (50.6933, 6.1291) & 191.33 / 101.33 \\
 R3 & (50.7200, 5.9210) & (50.5868, 5.9542) & (50.7409, 6.1309) & 170.99 / 80.99 \\
 \hline
 P1 & (46.0000, -125.0000) & (45.6513, -124.8726) & (46.0909, -124.5000) & 165.62 / 75.19 \\
 P2 & (46.0000, -125.0000) & (45.6403, -125.0170) & (45.9870, -124.4840) & 181.90 / 91.90 \\
 P3 & (46.0000, -125.0000) & (45.6491, -124.8855) & (46.0791, -124.4959) & 167.11 / 77.11 \\
 P4 & (46.0000, -125.0000) & (45.6580, -124.8399) & (46.1112, -124.5084) & 161.82 / 71.82 \\
 P5 & (46.0000, -125.0000) & (45.6403, -124.9819) & (46.0115, -124.4838) & 177.98 / 87.98 \\
 \hline
 Y1 & (29.0000, -94.0000) & (29.1798, -93.9827) & (29.0150, -94.2046) & 4.82 / 274.82 \\
 Y2 & (29.0000, -94.0000) & (29.0577, -93.8055) & (29.1709, -94.0660) & 71.29 / 341.29 \\
 Y3 & (29.0000, -94.0000) & (29.0265, -93.7969) & (29.1785, -94.0303) & 81.52 / 351.52 \\
 Y4 & (29.0000, -94.0000) & (28.9962, -93.7948) & (29.1804,  -93.9959) & 91.15 / 1.15 \\
 Y5 & (29.0000, -94.0000) & (29.0614, -93.8069) & (29.1696, -94.0701) & 70.06 / 340.06 \\
 \hline
\end{tabular}
\caption{Network sites coordinates and azimuths (clockwise from North). These are the same as described in \cite{Ebersold:2024hgp}.}
\label{tab:network_coords}
\end{table*}

\section{Discussion on Einstein Telescope alone}\label{app:ET_alone}
We briefly assess the detection prospects of a parity violating GWB using ET alone, building on previous ET-only detector design and sensitivity studies \cite{Branchesi:2023mws}. 
For this purpose, we consider baselines that differ from those adopted in the main analysis. Specifically, we introduce the notation $\Delta'_{10}$ and $\Delta'_{15}$ to denote closed, fully symmetric triangular ET configurations with $10$ km and $15$ km arms, respectively. These correspond to the idealized case of a perfectly equilateral triangle formed by three co-located detectors, with each vertex separated by 60°. For comparison, we also include the configurations $\Delta_{10}$ and $\Delta_{15}$ already used in the main text and defined in \cite{Ebersold:2024hgp}. These differ in that the three detectors do not form an exactly closed triangle. As a result, the $\Delta_{10}$ and $\Delta_{15}$ configurations are slightly asymmetric. 
This asymmetry yields enhanced sensitivity to $V$-modes, since the chiral ORF is slightly larger than that of a perfectly closed triangle.

In addition to triangular configurations, we also examine two 2L configurations, denoted by (SR)$_0$ and (SR)$_{45}$. The (SR)$_0$ setup corresponds to the optimal 2L configuration for GWB detection where the two detectors are aligned (relative orientation of $0$°). By contrast, the (SR)$_{45}$ represents the most pessimistic case for GWB detection, where the detectors are rotated by $45$° with respect to one another, consistent with previous studies of 2L configurations and relative orientations \cite{Branchesi:2023mws}. It is important to emphasize that (SR)$_0$ and (SR)$_{45}$ differ from (SR)$_1$ and (SR)$_2$ considered in the main analysis, since the latter were defined relative to the optimal configuration of a global network including two CEs, whereas here the reference is ET alone.
In Table \ref{tab:et_coords}, the latitude and longitude of each network center, the X- and Y-arm endpoints and azimuths are given.

\begin{table*}[htbp]
\centering
\footnotesize
\begin{tabular}{|c|c|c|c|c|}
 \hline
 \textbf{Name} & \textbf{Origin (lat, lon)} & \textbf{X-arm end (lat, lon)} & \textbf{Y-arm end (lat, lon)} & \textbf{Azimuth [°] X-/Y-arm} \\
 \hline
 E1'$_{10}$ & 43.6300, 10.5000 & 43.7149, 10.5413 & 43.6983, 10.4193 & 19.43 /319.43 \\ 
 E2'$_{10}$ & 43.7149, 10.5413 & 43.6983, 10.4193 & 43.6300, 10.5000 & 259.46 /199.46 \\
 E3'$_{10}$ & 43.6983, 10.4193 & 43.6300, 10.5000 & 43.7149, 10.5413 & 139.37 /79.37 \\
 \hline
 E1'$_{15}$ & 43.6300, 10.5000 & 43.7573, 10.5620 & 43.7325, 10.3789 &  19.43 /319.43 \\ 
 E2'$_{15}$ & 43.7573, 10.5620 & 43.7325, 10.3789 & 43.6300, 10.5000 & 259.47 /199.47 \\
 E3'$_{15}$ & 43.7325, 10.3789 & 43.6300, 10.5000  & 43.75730, 10.5620 & 139.35/79.35 \\
 \hline
 S$_0$ & 40.5200, 9.4170 & 40.3851, 9.4093  & 40.5258, 9.2402 & 182.51/272.51 \\
 S$_{45}$ & 40.5200, 9.4170 & 40.4203, 9.5364 & 40.4287, 9.2867 & 137.51 /227.51 \\
 \hline
 R$_0$=R$_{45}$ & 50.7200, 5.9210 & 50.8548, 5.9210  & 50.7198, 6.1334 & 0.00/90.00 \\
 \hline
\end{tabular}
\caption{ET coordinates and azimuths (clockwise from North).}
\label{tab:et_coords}
\end{table*}

\begin{figure}[htbp]
    \centering
    \includegraphics[width=\linewidth]{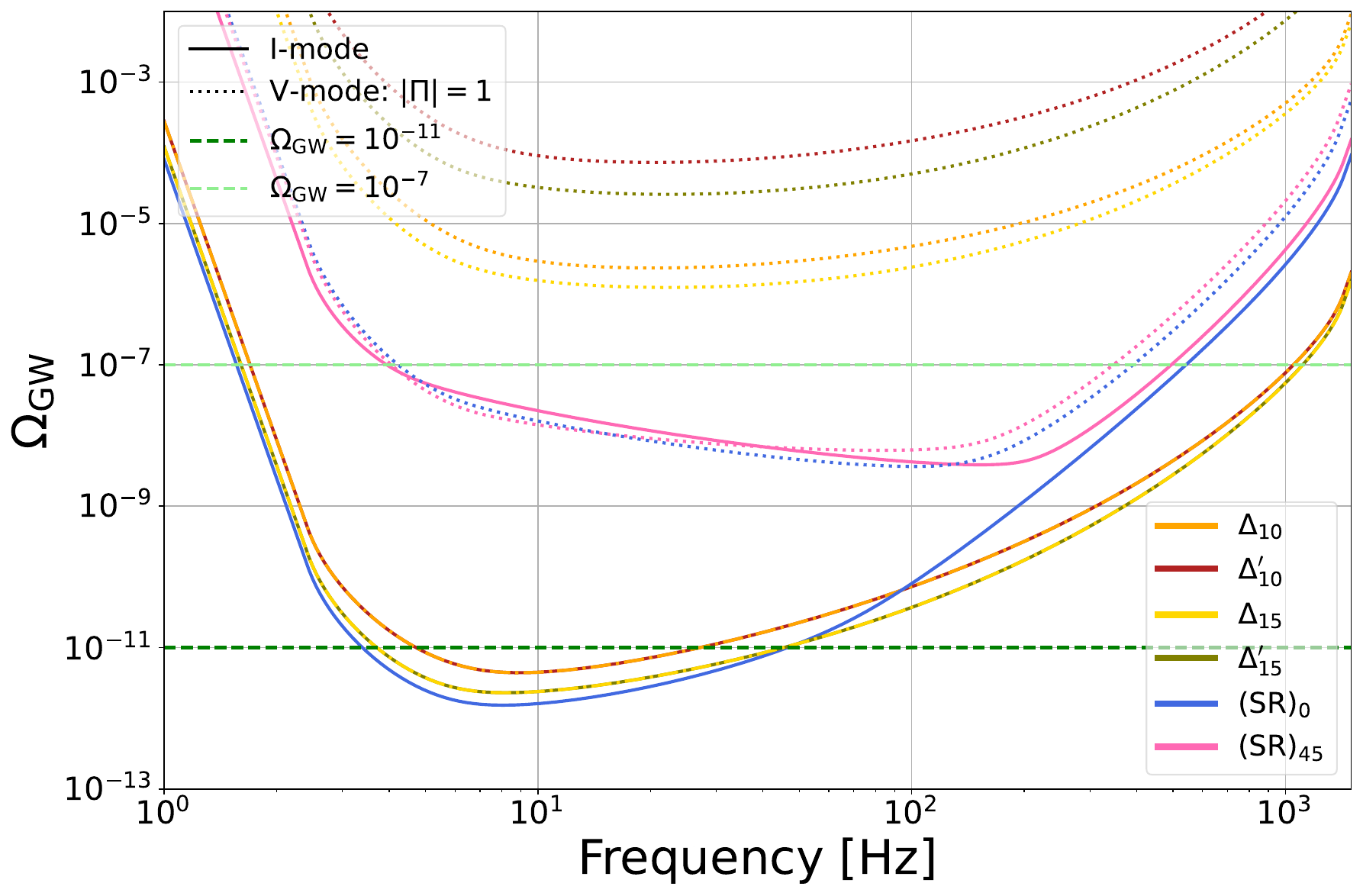}
    \caption{PI sensitivity curves for pure $I$-mode (solid lines) and pure $V$-mode GWBs (dotted and dashed lines), shown for the 6 different ET configurations, for an observing time of one year and $\rho=6$. The $V$-mode curves correspond to $|\Pi| = 1$ (dotted) and $|\Pi| = 0.1$ (dashed). We also show PL GWBs with spectral index of $\alpha=0$.}
    \label{fig: PI_ET_curves}
\end{figure}

\begin{figure}[htbp]
    \centering
    \includegraphics[width=\linewidth]{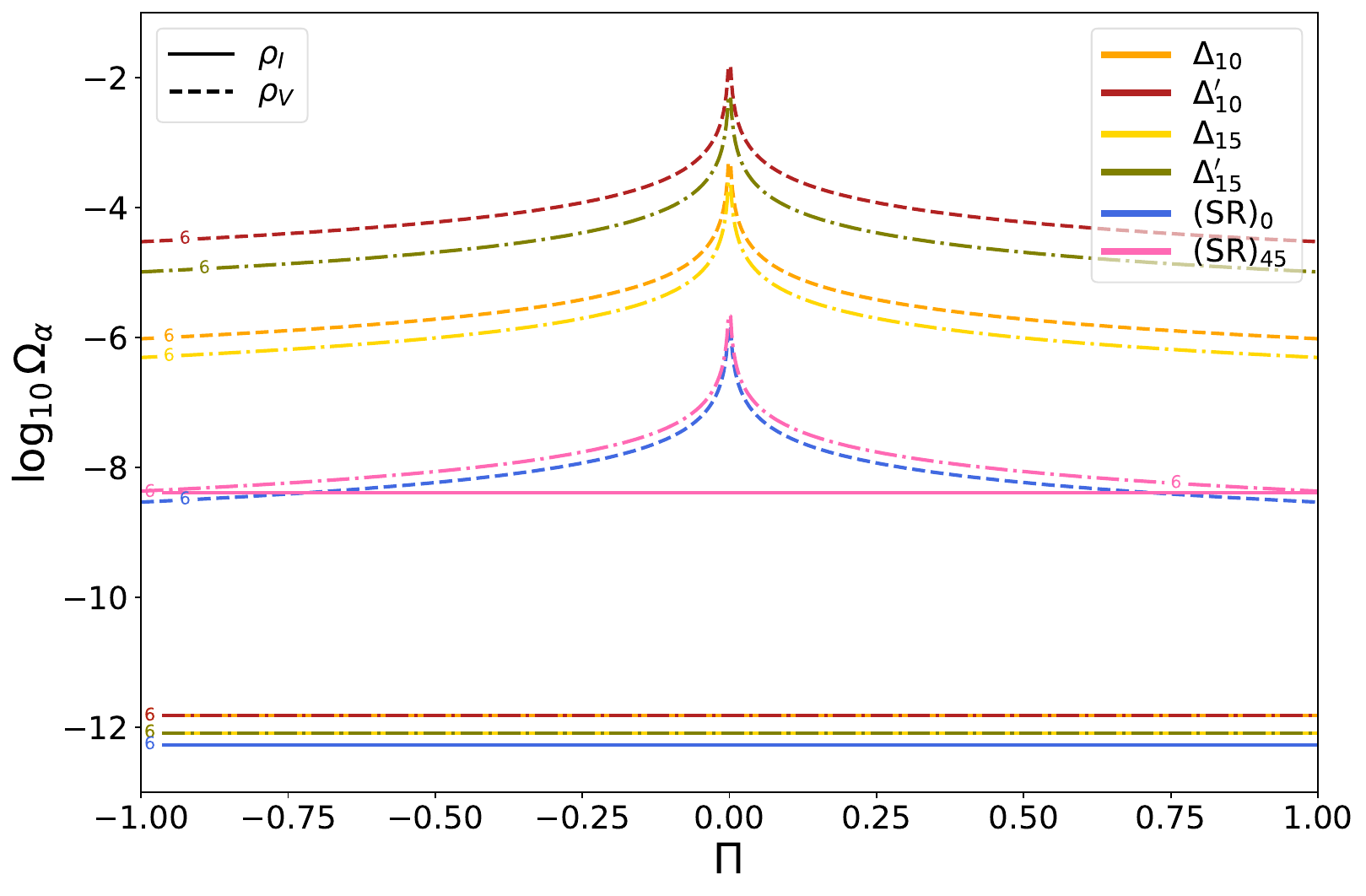}
    \caption{SNRs for pure $I$-mode (solid lines) and pure $V$-mode (dashed lines) GWBs, shown for the five different ET configurations for a PL GWB with $\alpha=0$. }
    \label{fig: SNR_ET_theory}
\end{figure}

Fig.~\ref{fig: PI_ET_curves} shows the PI curves for both the $I$- and $V$-modes across the different ET-only configurations. For the $I$-mode sensitivities, the trends are consistent with previous ET-only sensitivity studies \cite{Branchesi:2023mws}, while here we additionally present the corresponding $V$-mode sensitivities relevant for parity-violating GWBs. The triangular configurations are composed of three baselines and would be able to separate mixed $I$- and $V$-mode backgrounds, however since we are comparing them with networks with a single baseline, we assume pure $I$-and $V$-mode backgrounds in this case \cite{Seto:2007tn}. We note that for these we evaluated the ORF numerically to keep our results comparable to \cite{Caporali:2025dyf}.
For comparison, we also include a parity-violating PL GWB with spectral index $\alpha = 0$, to facilitate direct comparison with the corresponding SNR prospects shown in Fig.~\ref{fig: SNR_ET_theory}.
The perfectly closed and symmetric triangular configurations $\Delta'$ exhibit noticeably lower $V$-mode sensitivity than the slightly asymmetric $\Delta$ setups used in the main text, illustrating that even small asymmetries in ET’s geometry can enhance parity-violation detection. Nevertheless, all triangular configurations, whether symmetric or slightly asymmetric, are significantly less sensitive to the $V$-modes than the 2L designs, this is expected since the co-planar triangular ET detectors are weak to $V$-modes~\cite{Domcke_2020, Badger:2021enh, Seto:2008sr}.

\begin{table*}[htbp]
\begin{tabular}{|l|c|c|}
\hline
\diagbox
  {Rank}
  {\raisebox{-1.5ex}{$\alpha$}}  & $-3, -2, -1$ & $0, \tfrac{1}{2}, 1, \tfrac{2}{3}, 2, 3$ \\
\hline
1 & (SR)$_{45}$ &  (SR)$_{0}$\\
\hline
2 & (SR)$_{0}$ &  (SR)$_{45}$ \\
\hline
3 & $\Delta_{15}$ & $\Delta_{15}$ \\
\hline
4 & $\Delta_{10}$ & $\Delta_{10}$ \\
\hline
5 & $\Delta_{15}'$ & $\Delta_{15}'$ \\
\hline
6 & $\Delta_{10}'$ & $\Delta_{10}'$ \\
\hline
\end{tabular}
\centering
\caption{Ranking of ET configurations from most (1st) to least (6th) sensitive for different spectral indices $\alpha$ for the $V$-mode SNRs.}
\label{tab: SNR_V_rankings_ET}
\end{table*}

We have again performed SNR studies for multiple spectral indices $\alpha$.
For the $I$-mode SNRs, the ranking from most to least sensitive is: (SR)$_0$,   $\Delta_{15}=\Delta_{15}'$, $\Delta_{10}=\Delta_{10}'$ and (SR)$_{45}$. For the $V$-mode SNRs, the resulting network rankings are summarized in Tables~\ref{tab: SNR_V_rankings_ET}, confirming that 2L configurations consistently outperform triangular ET for $V$-mode detection. 
We note that correlated Newtonian \cite{Janssens:2022xmo, Janssens:2024jln, Caporali:2025mum} and magnetic \cite{Janssens:2021cta} noise between the co-located detectors of the triangular configuration is not included in the present study. Accounting for this contribution would further reduce the sensitivity prospects of the triangular ET designs.
Moreover, current bounds on the parity-violated GWB~\cite{LIGOScientific:2025kry} have ruled out the parity-violating parameter space that is detectable by both 2L and the $\Delta$ configurations, as seen in Figs.~\ref{fig: PI_ET_curves} and \ref{fig: SNR_ET_theory}. Consequently, if ET were constructed without CE, none of the configurations would be able to detect circularly polarized GWBs at a confident SNR level.

\clearpage
\bibliography{PV}

\end{document}